\documentstyle[preprint,aps]{revtex}
\begin{document}
\preprint{SNUTP 99-015; KIAS-P99014}
\draft
\title{Synchronization and resonance\\ 
in a driven system of coupled oscillators}

\author{H. Hong$^{1,3}$, M. Y. Choi$^{2,3}$, K. Park$^{2,3,*}$, B.-G. Yoon$^{4}$, 
and K.-S. Soh$^{1,3}$}
\address{$^1$Department of Physics Education, Seoul National University, 
Seoul 151-742, Korea}
\address{$^2$Department of Physics,
Seoul National University, Seoul 151-742, Korea}
\address{$^3$Center for Theoretical Physics,
Seoul National University, Seoul 151-742, Korea}
\address{$^4$Department of Physics, University of Ulsan, Ulsan 680-749, Korea}

\maketitle
\thispagestyle{empty}

\begin{abstract}
We study the noise effects in a driven system of  
globally coupled oscillators, with particular attention to
the interplay between driving and noise.
The self-consistency equation for the order parameter, which measures 
the collective synchronization of the system, is derived; it is found 
that the total order parameter decreases monotonically with noise,
indicating overall suppression of synchronization.
Still, for large coupling strengths, there exists an optimal noise level at which 
the periodic (ac) component of the order parameter reaches its maximum.
The response of the phase velocity is also examined and found to display 
resonance behavior.
\end{abstract}
\bigskip
\pagebreak

 
\section{Introduction}
\setcounter{equation}{0}

The set of coupled nonlinear oscillators serves as a prototype model
for a variety of self-organizing systems in physics and in other sciences, which
display the remarkable phenomena of collective 
synchronization~\cite{Winfree,Walker69,Eckhorn88,Benz90}.
Due to analytic simplicity and some physical as well as biological applications,
the system with global coupling has been mostly studied, 
both analytically and numerically~\cite{Kuramoto,Daido,Stro89,Arenas,Choi}.
Here external periodic driving may induce characteristic mode locking of each oscillator, 
leading the system to display periodic synchronization~\cite{MYChoi94}.
In such a driven system, the presence of noise raises another interesting
possibility of {\it stochastic resonance}~(SR),
which lead to amplification of the response of the system by 
cooperative interactions between the noise and external 
periodic driving~\cite{SR:rev}.
The SR phenomena, which have various practical 
applications~\cite{Fauve83,McNamara89,Fronzoni93,Collins95,Lindner95,Zhou90},
have been investigated in systems with relatively few degrees of freedom,
and observed in bistable systems
and also in systems with periodic potentials~\cite{Fronzoni93}.
On the other hand, the SR effects have hardly been examined in a system
with many degrees of freedom such as the system of coupled oscillators~\cite{Hakim}.

In this paper we consider a system of globally coupled stochastic oscillators, 
driven periodically, and investigate the interplay of noise and 
periodic driving, with particular attention to the possibility of stochastic 
resonance. 
For this purpose, it is crucial to consider appropriate responses of the 
system to the periodic forcing.
Here we consider the response of the phase velocity as well as
the order parameter which describes the phase synchronization.
We first derive the self-consistency equation for the 
order parameter and investigate the behavior of the order parameter
in the presence of noise.
It is found that the total order parameter, which consists of the time-independent (dc)
and periodic (ac) components, decreases monotonically with noise, indicating the
overall suppression of phase synchronization.
The ac component, on the other hand, may first increase as noise grows from zero,
and reach its maximum at a finite noise level.  Such SR-like behavior is also
observed in the response of the phase velocity:
At low noise levels, the noise subtracted power spectrum of the phase velocity  
tends to increase with noise.

This paper consists of six sections: Section II introduces the 
driven system of coupled oscillators subject to random noise.
The recurrence relation for the Fourier components is obtained.
In Sec.~III, we use the recurrence relation obtained in Sec.~II, and 
derive the self-consistency equation for the order parameter.
The set of coupled equations of motion for the system is transformed 
into a Fokker-Planck equation and the corresponding probability density 
is expanded as a Fourier series.
Sections IV and V are devoted to the investigation of the responses of the 
phase and of the phase velocity, respectively, to the external driving.  
In spite of the overall suppression of synchronization, 
the ac component of the order parameter,
corresponding to the phase response, as well as
the response of the phase velocity is revealed to display SR-like behavior.
Finally, a brief summary is given in Sec.~VI. 

\section{Driven System of Coupled Oscillators}

The set of equations of motion governing the dynamics of the 
system of $N$ coupled oscillators is given by
\begin{equation} \label{model}
 \dot{\phi_i} + {K\over N}\sum_{j=1}^N \sin(\phi_i-\phi_j)=
 \omega_i + I_i \cos\Omega t + \Gamma_i (t)  ~~~(i=1,2,...,N),
\end{equation}
where $\phi_i$ represents the phase of the $i$th oscillator.
The second term on the left-hand side corresponds to the global coupling 
between oscillators, with strength $K/N$.
The first and the second terms on the right-hand side describe the natural
frequency of the {\it i}th oscillator and the periodic driving on the 
{\it i}th oscillator, respectively.
Finally, $\Gamma_i (t)$ is independent white noise with zero mean and 
correlation
\begin{equation} \label{noiseproperty} 
\langle \Gamma_i (t)\Gamma_j (t')\rangle = 2D\delta_{i j}\delta(t-t'),
\end{equation}
where $D\,(>0)$ plays the role of the ``effective temperature'' of the system.
The natural frequency $\omega_i$ is distributed over the whole 
oscillators according to the distribution $g(\omega)$, which is assumed to be 
smooth and symmetric about $\omega_0$.  Without loss of generality, 
we may take $\omega_0$ to be zero and assume that $g(\omega)$ is 
concave at $\omega=0$, i.e., $g''(0)<0$.
The periodic (ac) driving amplitude $I_i $ may also vary for different 
oscillators, while the frequency $\Omega$ of the driving is assumed to 
be uniform for all oscillators. 
In the absence of noise ($D=0$), Eq.~(\ref{model}) precisely 
reduces to the set of equations of motion studied in Ref.~\cite{MYChoi94}.
The set of equations of motion in Eq.~(\ref{model}) describes 
a superconducting wire network~\cite{Park} and may also be regarded as 
the mean-field version of an array of resistively shunted junctions, 
which serves as a common model for describing the dynamics of 
superconducting arrays~\cite{Chung}.
In these cases, the two terms on the right-hand side of Eq.~(\ref{model}) 
correspond to the combined direct and alternating current bias.

Collective behavior of such an $N$-oscillator system is
conveniently described by the complex {\em order parameter}
\begin{equation} \label{deforder}
  \Psi \equiv {1\over N}\sum_{j=1}^N e^{i\phi_j} 
       = \Delta e^{i\theta},
\end{equation}
where nonvanishing $\Psi$ indicates emergence of synchronization.
Note that the synchronized state corresponds to the superconducting
state with global phase coherence in the case of the superconducting 
network or array.
The order parameter defined in Eq.~(\ref{deforder}) allows us to reduce
Eq.~(\ref{model}) into a {\em single} decoupled equation
$$
\dot{\phi_i} + K\Delta\sin(\phi_i-\theta)= \omega_i + I_i\cos\Omega t 
+ \Gamma_{i} (t).
$$
We then seek the stationary solution with $\theta$ being constant, which is possible
due to the symmetry of the distribution of $\omega_i$ and $I_i$ about zero.
Redefining $\phi_i{-}\theta$ as $\phi_i$ and suppressing
indices, we obtain the reduced equation of motion
\begin{equation} \label{single}
\dot{\phi} + K\Delta\sin\phi = \omega + I\cos\Omega t + \Gamma(t),
\end{equation}
which depends explicitly on the order parameter.
In this manner the order parameter $\Delta$, defined in terms of the phase 
via Eq.~(\ref{deforder}), in turn determines the behavior of the 
phase via Eq.~(\ref{single}), and
can thus be obtained by imposing self-consistency, as discussed in Sec.~III.
Note here that $\Delta$ in general depends periodically on time due to the 
periodic driving; this allows the Fourier expansion 
\begin{equation} \label{Deltaexpand}
\Delta=\Delta_0 + \sum_{s=1}^{\infty}\Delta_{s} \cos(s\Omega t + \alpha_s )
\end{equation}
with appropriate phases $\alpha_s$, where $\Delta_0$ is the time-independent 
(dc) component and $\Delta_s$ is the time-dependent (ac) one due to the periodic forcing.

A convenient way to deal with a set of Langevin equations is to introduce
an appropriate probability density and to resort to the associated
Fokker-Planck equation~\cite{Risken}. 
In general the set of $N$ Langevin equations (\ref{model}) makes it necessary
to consider the $N$-oscillator probability density $P(\{\phi_i\},t)$ and
the corresponding Fokker-Planck equation~\cite{Choi}.
In the system with glabal coupling, however, the set in Eq.~(\ref{model}) 
naturally reduces to the single Langevin equation~(\ref{single}), as shown above. 
This in turn leads to the Fokker-Planck equation for the single-oscillator
probability density $P(\phi, t)$ with the self-consistency for the order parameter
explicitly imposed, which has been considered in the absence of driving~\cite{Arenas}. 

The Fokker-Planck equation for the probability density $P(\phi,t)$ reads~\cite{Risken}:
\begin{equation} \label{Fokker}
\frac{\partial P}{\partial t} = 
\frac{\partial}{\partial\phi}
\left[\left(\frac{\partial V(\phi)}{\partial \phi}-I\cos\Omega t \right)P 
\right]
+D\frac{{\partial}^{2}P}{\partial\phi^2},
\end{equation}
where $V(\phi) \equiv - K\Delta\cos\phi - \omega\phi$ is the washboard potential.
Unlike the system without driving ($I=0$), the stationary solution of which 
has been obtained~\cite{Arenas},
Eq.~(\ref{Fokker}) does not allow such a simple stationary solution. 
We thus use the periodicity of the system and expand the probability 
density as a Fourier series 
\begin{equation} \label{probexpand}
P(\phi, t) = \sum_{n=-\infty}^{\infty} C_n (t)e^{i n\phi},
\end{equation}
which, upon substitution into Eq.~(\ref{Fokker}), yields 
\begin{eqnarray} \label{recurrence}
\dot C_n(t) &&= -\left[in(\omega + I\cos\Omega t) + n^2 D\right]
C_n (t) \nonumber \\
&&~~~~- \frac{n}{2} K\Delta (t) C_{n+1}(t)+\frac{n}{2}K\Delta (t)C_{n-1}(t).
\end{eqnarray}
Since the probability density should be real,
we have the relation $C_n = C^{*}_{-n}$; 
the normalization condition $\int_0^{2\pi}P(\phi,t) d\phi = 1$ 
gives the constant term $C_0 = 1 / 2\pi$.
The differential recurrence relation in Eq.~(\ref{recurrence})  
can be written in the form of an integral recurrence equation
\begin{eqnarray} \label{formalsol}
C_n(t) &&= C_n(0)
~\mbox{exp}\left[-in(\omega t+\frac{I}{\Omega}\sin\Omega t)-n^2 Dt\right]
\nonumber\\
&&~~~-\frac{n}{2}K ~ 
\mbox{exp} \left[-in(\omega t+\frac{I}{\Omega}\sin\Omega t)-n^2 Dt\right]
\nonumber\\
&&~~~~~\times \int_{0}^{t} dt' 
\Delta(t')
\biggl[ C_{n+1}(t')-C_{n-1}(t')\biggr]
~\mbox{exp}\left[in(\omega t'+\frac{I}{\Omega}\sin\Omega t')+n^2 Dt'\right],
\end{eqnarray}
which is of the same form as the equation for a single 
oscillator~\cite{Schleich84}, except for that 
here self-consistency for the order parameter is required.

\section{Self-consistency equation for the order parameter}

In this section we derive the self-consistency equation for the 
order parameter, which describes the response of the phase and 
determines the collective behavior of the system.
We suppose that the periodic driving amplitude $I$ is distributed according to
$f(I)$, independently of the natural frequency $\omega$.
Recalling that $\phi$ in Eq.~(\ref{single}) in fact represents $\phi-\theta$,
we have the self-consistency equation
\begin{eqnarray} \label{selfdelta}
  \Delta &=& {1\over N}\sum_j e^{i\phi_j} \nonumber \\
         &=& \int_{-\infty}^{\infty}dI\, f(I)
             \int_{-\infty}^{\infty}d\omega\, g(\omega)\,
             \langle e^{i\phi}\rangle_{\omega,I},
\end{eqnarray}
where $\langle \cdot \cdot \cdot \rangle_{\omega,I}$ denotes the average
with given $\omega$ and $I$.

With the probability density $P(\phi,t)$, the expansion of which is given 
by Eq.~(\ref{probexpand}), we compute the average 
\begin{equation} \label{exp}
\langle e^{i\phi} \rangle \equiv \int_{0}^{2\pi} d\phi~e^{i\phi}P(\phi, t)
= 2\pi C_1^{*}(t),
\end{equation}
where the relation $C_n = C_{-n}^*$ has been used.
Then Eq.~(\ref{selfdelta}) leads straightforwardly to 
\begin{equation} \label{sce}
\Delta = 2\pi \int_{-\infty}^{\infty} dI\, f(I) \int_{-\infty}^{\infty} 
d\omega \,g(\omega)C_1^{*}(t),
\end{equation}
which gives the order parameter in terms of the Fourier coefficient $C_1(t)$.
Assuming $K\Delta \ll 1$ near the transition to the coherent state,
we need to obtain $C_1$ up to the order of $(K\Delta)^3$.
For this, we first compute $C_2$ from 
Eq.~(\ref{formalsol}), neglecting $C_3$, and substitute the obtained $C_2$ 
back into the equation for $C_1$, i.e., Eq.~(\ref{formalsol}) with $n=1$.
At long times the transient terms such as $\mbox{exp} [-n^2 D t]$
for $n\neq 0$ vanish, and a lengthy calculation yields 
\begin{eqnarray} \label{C1}
C_1 (t)&&= -\frac{i K}{8\pi} \sum_{s=0}^{\infty}\Delta_s
\sum_{\ell,m} J_{\ell}(x) J_{m}(x) 
 \left[\frac{e^{i(\ell-m+s)\Omega t +i\alpha_s}}{\omega+(\ell+s)\Omega -iD}
      +\frac{e^{i(\ell-m-s)\Omega t -i\alpha_s}}{\omega+(\ell-s)\Omega -iD}
 \right] \nonumber\\
&&~~-\frac{i K^3}{64\pi} 
\sum_{s,s',s''} \Delta_{s} \Delta_{s'} \Delta_{s''}
\sum_{\ell, \ell', m, m', n, n'}
J_{\ell}(x)J_{\ell'}(x)J_{m}(2x)J_{m'}(2x)J_{n}(x)J_{n'}(x) \nonumber \\
&&~~~~~\times 
\left[
 e^{i(\alpha_s-\alpha_{s'}-\alpha_{s''})} F(t; s,s',s'') 
+e^{i(\alpha_s+\alpha_{s'}-\alpha_{s''})} F(t; s,-s',s'') 
\right. \nonumber\\
&&~~~~~~+e^{i(\alpha_s-\alpha_{s'}+\alpha_{s''})} F(t; -s,s',s) 
+e^{i(\alpha_s+\alpha_{s'}+\alpha_{s''})} F(t; -s,-s',s) \nonumber\\
&&~~~~~~
+e^{i(-\alpha_s-\alpha_{s'}-\alpha_{s''})} F(t; s,s',-s) 
+e^{i(-\alpha_s+\alpha_{s'}-\alpha_{s''})} F(t; s,-s',-s) \nonumber \\
&&~~~~~~\left.
+e^{i(-\alpha_s-\alpha_{s'}+\alpha_{s''})} F(t; -s,s',-s'') 
+e^{i(-\alpha_s-\alpha_{s'}-\alpha_{s''})} F(t; -s,-s',-s'') 
\right],
\end{eqnarray}
where $x\equiv I/\Omega$, $\alpha_0 \equiv 0$, and
the function $F(t; s,s',s'')$ depends on the indices $\ell, \ell', m, m', n$, and $n'$
as well as on $\omega$ and $\Omega$:
\begin{eqnarray*}
F(t; s,s',s'') &\equiv&
  \frac{e^{i(\ell-\ell'+m-m'+n-n'-s-s'-s'')\Omega t}}
       {[\omega+(n'+s)\Omega -iD][2\omega+(m'-n+n'+s+s')\Omega -4iD]} \\
       & & \times \frac{1}{\omega+(\ell'-m+m'-n+n'+s+s'+s'')\Omega -iD}.
\end{eqnarray*}
With the above expression for $C_1$, Eq.~(\ref{sce}) yields the explicit form of 
the self-consistency equation for the order parameter.

Comparing term by term in the resulting self-consistency equation, we can determine
each component of the order parameter.  Namely, the dc component
$\Delta_0$ is given by the constant (zero-frequency) terms in the expansion of $C_1(t)$.
The next component $\Delta_1$ can be obtained from the terms with frequency $\Omega$,
the component $\Delta_2$ from the $2\Omega$ terms, and so on.
For weak driving, the ac components of the order parameter are much smaller than the
dc component, leading to the simple self-consistency equation:
\begin{equation}\label{self}
\Delta \approx a K\Delta_0 - b(K\Delta_0)^3 
\end{equation}
with the coefficients
\begin{eqnarray} \label{coeff}
a &&=\frac{i}{2} \sum_{\ell,m} \int d I \,f(I) J_{\ell}(x)J_m(x) 
e^{i(\ell-m)\Omega t}
\int d\omega \,\frac{g(\omega)}{\omega+m\Omega+iD}
\nonumber \\
b &&= - \frac{i}{4} \sum_{\ell, \ell', m, m', n, n'}
\int dI\, f(I) 
J_{\ell}(x)J_{\ell'}(x)J_{m}(2x) J_{m'}(2x)J_{n}(x)J_{n'}(x) 
\,e^{i(\ell'-\ell+m'-m + n'-n)\Omega t} \nonumber\\
&&~~~\times 
\int d\omega \,\frac{g(\omega)}
{[\omega+n'\Omega+iD][2\omega+(m'+n'-n)\Omega+4iD]} \nonumber \\
&&~~~~~\times \frac{1}{\omega+(\ell'+m'-m+n'-n)\Omega+iD}.
\end{eqnarray}
In the simple case of no external driving ($I=0$) and noise ($D\rightarrow 0$),
the representation $\pi\delta(\omega)=D(\omega^2 +D^2)^{-1}$
in the limit $D\rightarrow 0$ together with the symmetry of $g(\omega)$
reduces Eq.~(\ref{coeff}) to $a=(\pi/2)g(0)$ and $b=-(\pi/16)g''(0)$,
which indeed reproduces the self-consistency equation obtained 
in Ref.~\cite{Kuramoto}.

Solving Eq.~(\ref{self}), we obtain the collective behavior of the system,
which has been analyzed in Ref.~\cite{MYChoi94}: 
For small $K$, only the trivial solution $\Delta=0$ exists.
On the other hand, for $K\geq K_c \equiv 1/a_0$,
Eq.~(\ref{self}) also allows the nontrivial solution
with the dc component
\begin{equation} \label{solu}
\Delta_0 = \Delta_{+} \equiv \frac{\sqrt{b_0 K(a_0 K-1)}}{b_0 K^2},
\end{equation}
where the constant coefficients are given by
\begin{eqnarray} \label{coeff0}
a_0 &&=\frac{i}{2} \sum_{\ell} \int d I \,f(I) J_{\ell}^2 (x) 
\int d\omega \,\frac{g(\omega)}{\omega+\ell\Omega+iD}
\nonumber \\
b_0 &&= - \frac{i}{4} \sum_{\ell, m, m', n, n'}
\int dI\, f(I) 
J_{\ell}(x)J_{m}(2x) J_{m'}(2x)J_{n}(x)J_{n'}(x)J_{\ell+m+n-m'-n'}(x) \nonumber\\
&&~~~~~\times 
\int d\omega \,\frac{g(\omega)}
{[\omega+n'\Omega+iD][\omega+\ell\Omega+iD][2\omega+(m'+n'-n)\Omega+4iD]}.
\end{eqnarray}
Thus as $K$ is increased beyond $K_c$, the null solution becomes unstable
and the stable nontrivial solution $\Delta_{+}$ (accompanied by the ac components)
appears via a pitchfork bifurcation at $K=K_c$~\cite{MYChoi94}.

\section{Noise Effects on Synchronization}

To understand the cooperative effects of the driving and noise 
on the response of the system, 
we examine in this section how the noise affects synchronization behavior of the system.
For simplicity, we consider the weak-driving or high-frequency limit ($x \equiv I/\Omega \ll 1$),
expand the Bessel functions in Eq.~(\ref{coeff}) to the order of $x^2$,
and perform the average over the distribution $f(I)$.
This gives the coefficient $a$ to the order of $\sigma_I$,
the variance of the distribution $f(I)$:
\begin{eqnarray}\label{a}
a &=& \frac{D}{2} \int d\omega\,\frac{g(\omega)}{\omega^2 + D^2}
   + \frac{\sigma_{I}}{8\Omega^2} \int d\omega \,g(\omega)\,
         \left[\frac{D(\cos 2\Omega t - 2)}{\omega^2 + D^2} 
         \right. \nonumber\\
& &~~~-\frac{D(\cos 2\Omega t-1)+(\omega+\Omega)\sin 2\Omega t}
           {(\omega+\Omega)^2+D^2} 
     -\frac{D(\cos 2\Omega t-1)-(\omega-\Omega)\sin 2\Omega t}
           {(\omega-\Omega)^2+D^2}  \nonumber \\
& &~~~\left.
     +\frac{D\cos 2\Omega t+(\omega+2\Omega)\sin 2\Omega t}
           {2(\omega+2\Omega)^2 + 2D^2}
     +\frac{D\cos 2\Omega t -(\omega-2\Omega)\sin 2\Omega t}
           {2(\omega-2\Omega)^2+2D^2} \right] \nonumber\\
 &\equiv& a_0 + a_2\cos(2\Omega t+\alpha_2).
\end{eqnarray}
Similarly, a tedious but straightforward calculation leads to the coefficient 
$b=b_0 + b_2\cos(2\Omega t +\alpha_2)$.  Note that the symmetry of the distribution
$f(I)$ about $I=0$ forbids the frequency $\Omega$ term, which is linear in the driving.
Here it is easy to observe that $a_0$, which reads
\begin{eqnarray}\label{a0}
a_0 &=& \frac{D}{2}\left(1-\frac{\sigma_I}{2\Omega^2}\right)
        \int d\omega\,\frac{g(\omega)}{\omega^2 + D^2}\nonumber \\
    & &~+ \frac{D\sigma_{I}}{8\Omega^2} \int d\omega \,g(\omega)
          \left[\frac{1}{(\omega+\Omega)^2+D^2}+\frac{1}{(\omega-\Omega)^2+D^2}
          \right],
\end{eqnarray}
in general decreases monotonically with $D$.
Thus the critical coupling strength $K_c$ grows as the noise level is raised.
Figure 1 displays the monotonic increase of $K_c \,(={a_0}^{-1})$ with the noise level $D$,
for $\sigma_{I}=0.1$ and $\Omega=2.0$.  
For the distribution of natural frequencies, the Gaussian distribution with 
variance $\sigma_{\omega} = 0.5$ has been chosen.

With the coefficients $a$ and $b$ obtained above, the order parameter can be obtained
from Eq.~(\ref{self}) and its behavior in the presence of noise can be investigated. 
Indeed the dc component $\Delta_0$ given by Eq.~(\ref{solu}) is easily found to decrease
monotonically as the noise level $D$ is increased.  
On the other hand, it is too complicated to obtain analytically the explicit 
behavior of the ac component $\Delta_s$ $(s \geq 1)$.
Further, the analytical results are based on 
Eq.~(\ref{self}), which is valid only near the transition ($K \approx K_c$);
this makes it desirable to obtain the order parameter numerically.
We have thus performed numerical simulations to compute the components 
$\Delta_0$ and $\Delta_2$.
Since the effects of external driving first appear in the coefficients $a_2$ and $b_2$,
giving rise to the frequency $2\Omega$ term,
it is relevant to investigate $\Delta_2$ as the response to the external driving.
In the simulations Eq.~(\ref{model}) has been integrated 
with discrete time steps of $\Delta t=0.01$. 
At each run, we have used $N_t=6048$ 
time steps to compute the order parameter, discarding 
the data from the first $4\times 10^3 $ steps, and varied both
$\Delta t$ and $N_t$ to verify that the stationary state was achieved.
Finally, independent runs with $30$ different distributions of the 
natural frequency and initial conditions have been performed,
over which the averages have been taken.
For both the distribution of the driving amplitudes and that of the natural frequencies,
we have chosen Gaussian distributions with various values 
of variances $\sigma_{I}$ and $\sigma_{\omega}$, only to find
no qualitative difference.

Figure 2 displays the obtained behaviors (a) of the dc component $\Delta_0$
and (b) of the ac component $\Delta_2$ 
in the system of $N=1000$ oscillators, driven by frequency $\Omega = 1.0738$
and with variances $\sigma_{\omega}=1.0$ and $\sigma_{I}=1.0$.
The data represented by empty and solid squares in Fig.~2 correspond to the 
coupling strength $K=2.5$ and $K=3.0$, respectively.
It is shown that for both values of the coupling strength, 
$\Delta_0$ decreases monotonically as the noise level $D$ is raised.
Such monotonic behavior is also exhibited by $\Delta_2$ for $K=2.5$.
For $K=3.0$, on the other hand, Fig.~2(b) displays that the ac component $\Delta_2$ 
first increases with noise and reach its maximum at a finite value of the noise level $D$.
Similar non-monotonic behavior of $\Delta_2$ can be observed 
for larger values of the coupling strength $K$, suggesting 
the presence of SR-like behavior in the order parameter.
Note, however, that the dc component $\Delta_0$ is in general dominant over the ac
component, leading to the monotonic decrease of the total order parameter.
It is thus concluded that noise tends to suppress monotonically the overall synchronization in 
the system.

\section{Response of the Phase Velocity}

In this section we investigate the power spectrum of the phase velocity
at the driving frequency,
which conveniently describes the response of the phase velocity to the external driving.
In the case of a superconducting wire network or array,
the phase velocity can be identified with the voltage via the Josephson relation,
and the power spectrum of the phase velocity simply corresponds to the voltage
power spectrum under the combined direct and alternating current driving.
Equations (\ref{single}) and (\ref{probexpand}) give
the average phase velocity of a single oscillator 
in terms of $\mbox{Im}C_1$, the imaginary part of $C_1$:
\begin{eqnarray} \label{phasevel}
\langle \dot\phi\rangle &\equiv& \int_{0}^{2\pi}  d\phi\, P(\phi,t)\dot\phi \nonumber\\
 &=& \omega + I\cos\Omega t + 2\pi K\Delta~\mbox{Im}C_1,
\end{eqnarray}
which, upon substitution of Eq.~(\ref{C1}) for Im$C_1$, obtains the simple form
\begin{equation}
\langle\dot\phi\rangle = \omega + A\cos\Omega t + B\sin\Omega t 
+ O(K^2 \Delta_0 \Delta_1 )
\end{equation}
with the amplitudes
\begin{eqnarray}
A &=& I \Biggr[ 1-\frac{K^2 {\Delta_0}^2 }{2\Omega}
\Biggl( \frac{\omega+\Omega}{(\omega+\Omega)^2 + D^2 }
-\frac{\omega-\Omega}{(\omega-\Omega)^2 +D^2 }\Biggr)\Biggr], \nonumber\\
B &=& \frac{K^2{\Delta_0}^2 }{2\Omega}DI
\Biggl[\frac{1}{(\omega+\Omega)^2+D^2 } - \frac{2}{\omega^2 + D^2 } 
+\frac{1}{(\omega-\Omega)^2 +D^2 }\Biggr]. \nonumber
\end{eqnarray}

The desired power spectrum $S$ of the phase velocity at the driving frequency
is proportional to the square of the Fourier component of frequency $\Omega$,
i.e., $S(\Omega) \propto A^2 + B^2$.
In the limit $D \rightarrow 0$, the amplitude of the Fourier component approaches
$$
I^2 \left[1+ \frac{K^2 \Delta_0 ^2}{\omega^2 - \Omega^2} \right]^2 
 + I^2 \frac{\pi^2 K^4 \Delta_0 ^4}{4\Omega^2}
 [\delta(\omega+\Omega)-2\delta(\omega)+\delta(\omega-\Omega)]^2 
$$
while it approaches $I^2$ in the limit $D\rightarrow \infty$.
It is of interest to note that the amplitude in the noiseless limit
can be either larger or smaller than that in the strong-noise limit,
depending on $\omega$ and $\Omega$:
As the noise level is raised, the amplitude tends to decrease from the noiseless value
for $\omega > \Omega$ and increase for $\omega < \Omega$.
Accordingly, for given driving frequency, those oscillators with smaller/larger natural
frequencies contribute to the increase/decrease of the amplitude toward its
asymptotic value $I^2$.
For small values of the variance $\sigma_{\omega}$, for example, 
most oscillators should possess the natural frequency
$\omega < \Omega $, although there may still exist some oscillators with frequency 
$\omega > \Omega $.  
The power spectrum of the whole system, which is given by the sum of
contributions from all the oscillators, is then
expected to increase for small $D$ and to approach the asymptotic value which is 
proportional to $I^2$.
Unfortunately, however, the approximations used on various stages of the
analysis disallows a reliable analysis. In particular
the extrapolation to the limit $D\rightarrow 0 $ is untrustworthy since 
nonzero effective temperature $(D\neq 0)$ has been assumed in solving the equation for
$P(\phi,t)$.
It is also obvious that the higher-order terms neglected in the analysis 
set a limit in the regime of validity, making it desirable to investigate
the system by other means.

We have thus performed numerical simulations to obtain the power spectrum
for various values of the coupling strength and of the variance in the distributions
of the natural frequency and of the driving amplitude.
We have again integrated Eq.~(\ref{model}) for the system of $N=1000$ oscillators
with discrete time steps of $\Delta t=0.01$, using
at each run $N_t = 6048$ time steps to compute 
the power spectrum of the phase velocity and discarding
the data from the first $4\times 10^3$ steps.
The averages have been taken over $300$ independent runs 
with different distributions of the natural frequency and 
initial conditions.
>From the obtained time series, we have computed the power spectrum 
by means of the fast Fourier transform algorithm.
To take into account the background noise,
we have taken five nearest data points around the peak at the driving frequency 
in the power spectrum and performed the average to give the noise level. 
(The results have been found not to change qualitatively even if other measure
for the noise level is adopted.)

In Fig. 3 we present the obtained data: the background noise subtracted 
power spectrum of the phase velocity at the driving frequency versus the noise level.
For the distributions of the natural frequency and of the driving amplitude,
Gaussian ones have been chosen with variances $\sigma_{\omega}=0.5$ and 
$\sigma_I=0.2$, respectively, while
the coupling strength $K=2.5$ and the driving frequency 
$\Omega = \pi/1.024$ have been taken.
Remarkably, Fig.~3 displays that the power spectrum increases as the noise level is raised from zero.
Obviously it does not keep increasing monotonically with the noise,
and there apparently exists an optimal noise level at which the power spectrum reaches 
its maximum.  
Beyond the optimal noise level, the power spectrum
first falls off gradually and saturates eventually toward its asymptotic value,
although this behavior is somewhat obscured by the large fluctuations 
due to strong noise.
Such a broad peak followed by gradual decrease 
has also been observed in the SR of another system~\cite{Collins95}.
It is thus suggested that the response of the phase velocity 
also displays SR-like behavior in the appropriate regime.
Since the phase velocity corresponds to the voltage in a superconducting system, 
this indicates that the noise subtracted power spectrum of the voltage
displays such resonance behavior.

\section{Summary}

We have studied the noise effects in a driven system of 
globally coupled oscillators, with emphasis on the interplay of noise and periodic driving.
In particular, to investigate the possibility of resonance behavior,
we have considered the response of the phase velocity as well as
the order parameter which describes the phase synchronization.
The self-consistency equation for the order parameter, derived from 
the recurrence relation of the probability density, has been shown to
display monotonic decrease of the total order parameter
in the presence of noise.  It has thus been concluded
that noise in general suppresses overall phase synchronization in the system,
i.e., superconductivity tends to be disturbed by noise present in a superconducting 
wire network or array.

Nevertheless it has also been revealed that for large coupling strengths
the ac component of the order parameter increases with the noise level growing from zero
and reach its maximum at a finite noise level.  
Such resonance behavior has also been observed in the response of the phase velocity:
At low noise levels, the noise subtracted power spectrum of the phase velocity
has been found to increase with noise, displaying a broad peak at a finite noise level.
As the noise level is raised further, the power spectrum
appears to saturate toward its asymptotic value,
although concealed by large fluctuations due to strong noise.
In conclusion, the phase synchronization, describing the collective behavior of the 
coupled-oscillator system, is suppressed monotonically in the presence of noise.
Still the responses of the phase and of the phase velocity 
can display non-monotonic resonance behavior in the appropriate regime,
which may be manifested by a broad resonance peak of the voltage power spectrum
in the case of a superconducting system.

It is also of interest to note that the phase velocity on average may serve as
a measure of $\dot{\Delta}/\Delta$, the rate of change of phase synchronization.
Accordingly, the resonance behavior in the response of the phase velocity
suggests that the approach to the coherent state with synchronization 
can be accelerated by the presence of weak noise.  
Since the phase synchronization corresponds to the memory retrieval 
in the network of neuronal oscillators~\cite{Arenas,Choi},
the resonance behavior may also imply the information processing assisted
by weak noise in a biological system, e.g., the crayfish who appears to
use such resonance to perceive an enemy quickly.
The detailed understanding of the resonance behavior 
and its implications to applicable physical and biological systems
require more extensive analytical and numerical investigations, 
which are left for further study.

\section*{Acknowledgments}

MYC thank C.~W. Kim for the hospitality during his stay at Korea Institute for Advanced Study, 
where part of this work was accomplished.
This work was supported in part by the Seoul National University Research Fund, 
by the Korea Research Foundation, and by the Korea Science and Engineering Foundation.
BGY also acknowledges the partial support from the Research Fund of University of Ulsan in 1999.

\pagebreak

\begin{figure}
\vspace*{15.5cm}
\includegraphics{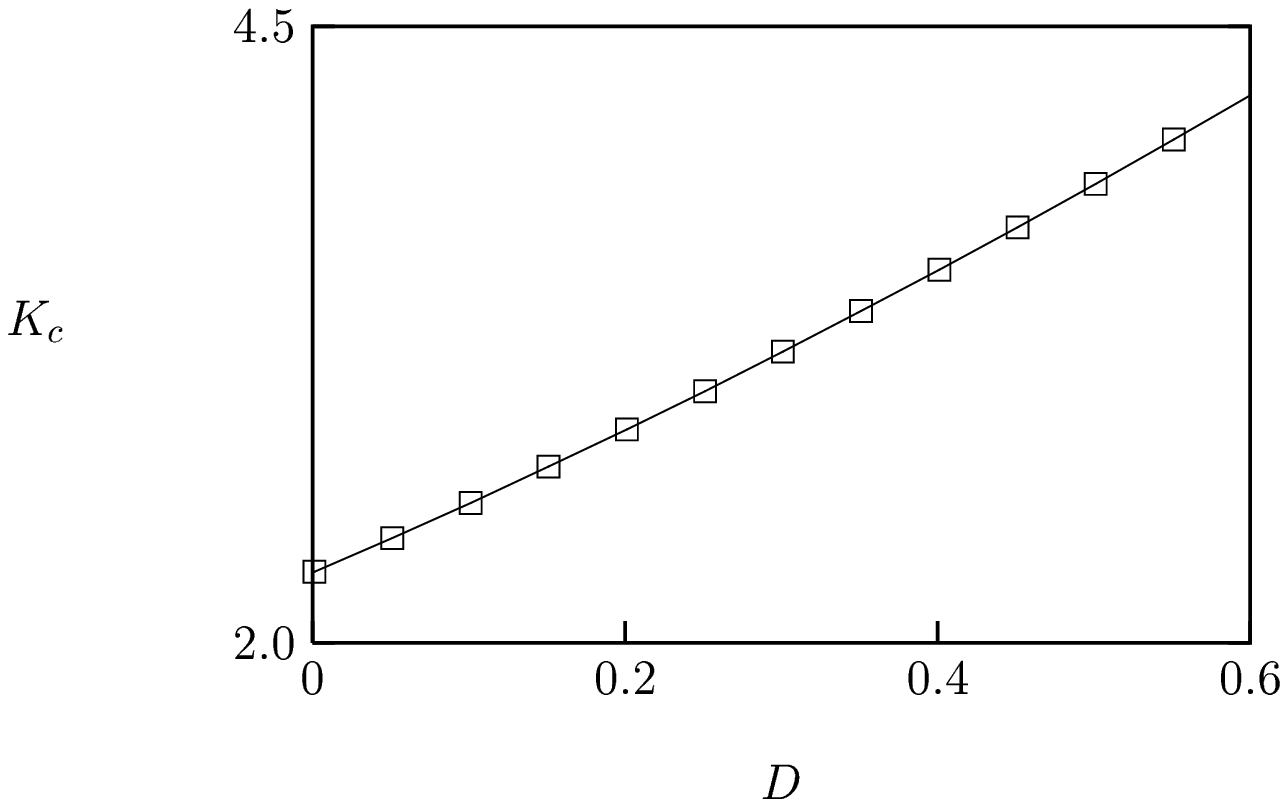}
\caption{
     The critical coupling strength beyond which synchronization appears versus 
     the noise level in the system with the driving frequency $\Omega=2$ and the variances
     $\sigma_{\omega}=0.5$ and $\sigma_{I}=0.1$. 
     The random noise in the system monotonically increases the critical coupling strength,
     thus tending to suppress synchronization.}
\end{figure}

\begin{figure}
\vspace*{15.5cm}
\includegraphics{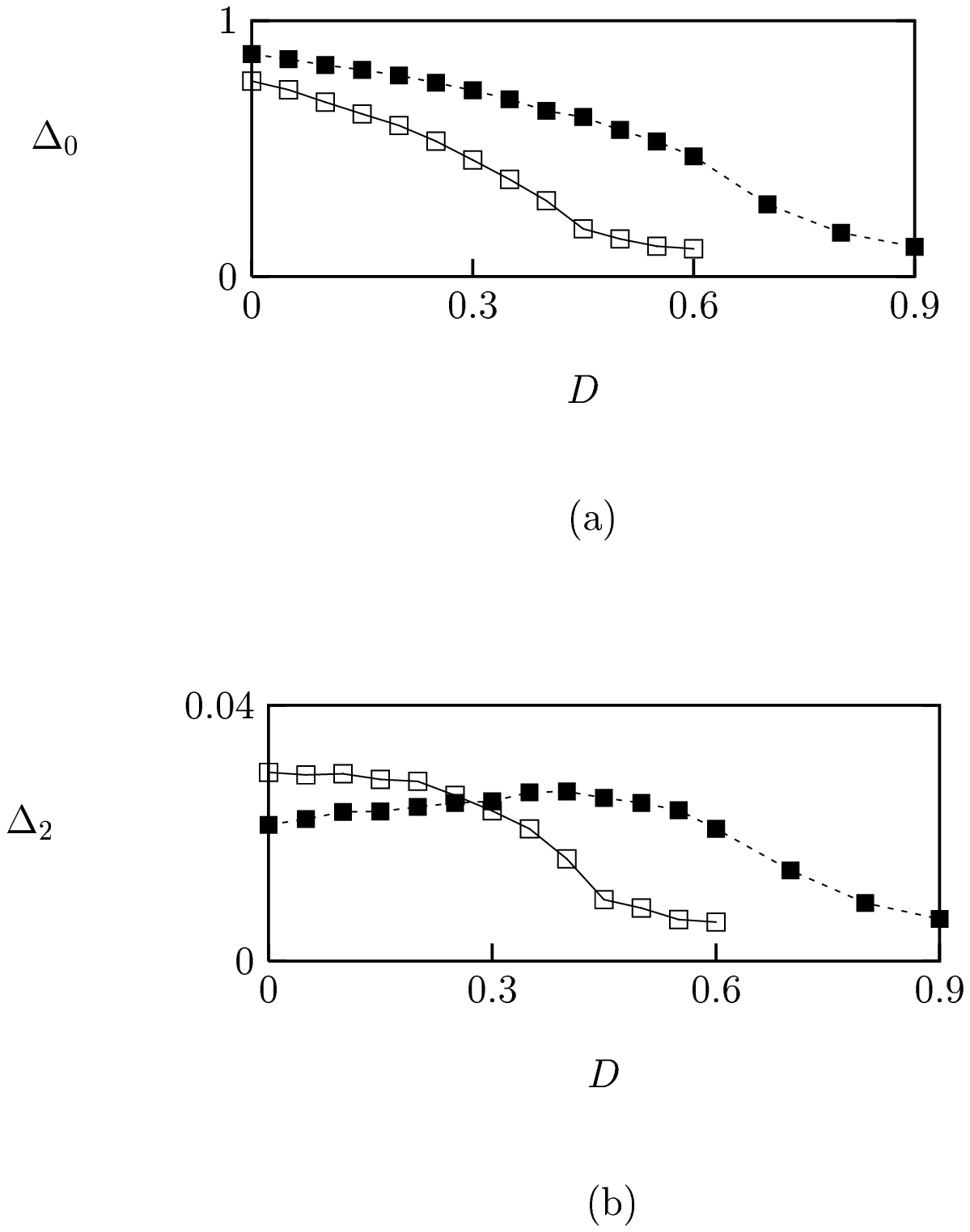}
\caption{
      Behavior of the order parameter in the presence of noise.
      The data represented by empty and solid squares correspond to the coupling strength
      $K=2.5$ and $K=3.0$, respectively.
      (a) The dc component $\Delta_0 $ for both $K=2.5$ and $K=3.0$ is shown to decrease 
      monotonically.
      (b) The ac component $\Delta_2 $ for $K=2.5$ decreases monotonically, while 
      for $K=3.0$ it displays a peak at a finite noise level.  The standard deviations of 
      the data (not shown) range from $5\%$ to $15\%$, and lines are
      merely guides to the eye.}
\end{figure}

\begin{figure}
\vspace*{15.5cm}
\includegraphics{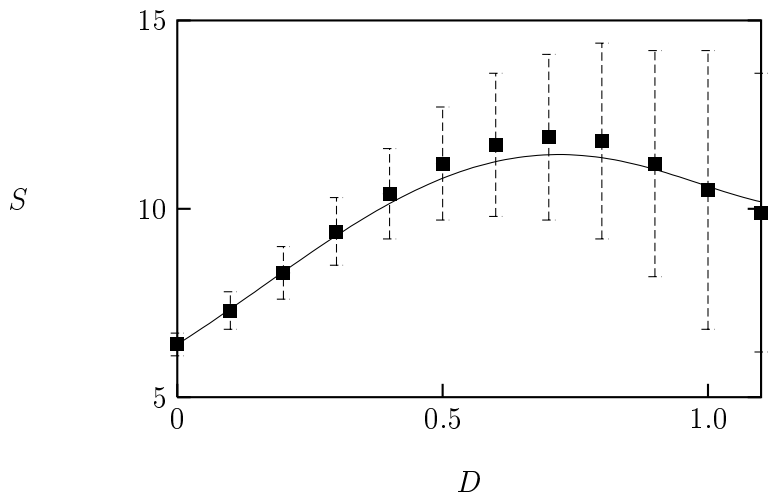}
\caption{ 
      The noise subtracted power spectrum of the phase velocity 
      at the driving frequency.  
      There appears an optimal noise level at which the 
      power spectrum reaches its maximum.  The error bars have been estimated by the 
      standard deviation and the line is merely a guide
      to the eye.}
\end{figure}
\end{document}